\documentclass[conference]{IEEEtran}
\usepackage[cmex10]{amsmath}
\usepackage{graphicx}
\usepackage{epstopdf}
\usepackage{cite}
\usepackage{algorithm} 
\usepackage{algorithmic} 
\usepackage[caption = false]{subfig}

\begin{document}
\title{User Association for Offloading in Heterogeneous Network Based on Matern Cluster Process}
\author{\IEEEauthorblockN{Yuxuan~Xie\IEEEauthorrefmark{1}, Xuefei~Zhang\IEEEauthorrefmark{1}, Qimei~Cui\IEEEauthorrefmark{1} and Yanyan~Lu\IEEEauthorrefmark{1}}
\IEEEauthorblockA{\IEEEauthorrefmark{1}Key Laboratory of Universal Wireless Communication, Ministry of Education\\
Beijing University of Posts and Telecommunications, Beijing, 100876, China
\\Email: \IEEEauthorrefmark{1}\{cuiqimei, zhangxuefei\}@bupt.edu.cn}
}
\maketitle

\begin{abstract}
Future mobile networks are converging toward heterogeneous multi-tier networks, where various classes of base stations (BS) are deployed based on user demand. So it is quite necessary to utilize the BSs resources rationally when BSs are sufficient. In this paper, we develop a more realistic model that fully considering the inter-tier dependence and the dependence between users and BSs, where the macro base stations (MBSs) are distributed according to a homogeneous Poisson point process (PPP) and the small base stations (SBSs) follows a Matern cluster process (MCP) whose parent points are located in the positions of the MBSs in order to offload the users from the over-loaded MBSs. We also assume the users are just randomly located in the circles centered at the MBSs. Under this model, we derive the association probability and the average ergodic rate by stochastic geometry. An interesting result that the density of MBS and the radius of the clusters jointly affect the association probabilities in a joint form is obtained. We also observe that using the clustered SBSs results in aggressive offloading compared with previous cellular networks.
\end{abstract}

\section{Introduction}\label{sec1}
The fact that wireless mobile networks are facing explosive data traffics, especially video streams,  pushes us to find complementary alternatives to ease the pressure of MBSs. Under this background, heterogeneous network came into being with the deployment of multiple classes of BSs that differ in terms of maximum transmit power, physical size, ease-of-deployment and cost [1]. The deployment of multi-class BSs can not only compensate for the coverage loopholes of the macroBSs, but also transfer the over-load traffic from MBSs to other low-power BSs, named cellular offloading, in order to relieve the macro BSs' service pressure coming from the increasing user demands [2]. From the perspective of users in such heterogeneous network, user association plays a pivotal role in cellular offloading and enhancing the load balancing, the spectrum efficiency, and the energy efficiency of networks [3][4][5].

Recently, many works have been done to analyze performance metrics (such as SINR distribution, the coverage/outage probability and average rate) in HetNet using the typical user methodology in stochastic geometry [6-10] in comparison with traditional cellular network. Further, researchers derive the association probability in HetNet, a key metric on offloading [6], [9] representing the probability that a typical user is associated with a certain tier. Specifically, literature [1], [6], [9] derived different performance metrics (e.g., the coverage probability, average rate) under their respective system model. There are subtle differences among these models, but they all assume that the locations of BS follow a homogeneous Poisson point process(PPP) for single-tier network, or multiple tiers of mutually independent PPPs for heterogeneous cellular networks (HCNs). In addiction, independent of the BSs' deployment, the users distribution also follow HPPP. Practically, human activities are hardly completely random and trend to be clustered. Although the assumption of PPP makes the analysis tractable, it dose not seem realistic in the case of non-uniform user distributions. And the network operators trend to deploy the SBSs at where more people aggregate (in order to offload the pressure of MBS), we expect that the locations of SBSs to be clustered. Several models of cluster process are described in detail in [8]. Poisson cluster processes (PCP) result from homogeneous independent clustering applied to a homogeneous Poisson process. The parent points form a homogeneous Poisson process while the daughter points of a representative cluster are random in number and are scattered independently with identical spatial probability density around the origin. We further focus on one of more specific models for the representative cluster, namely Matern cluster processes (MCP). In MCP, the number of points in the representative cluster has a Poisson distribution with the mean  $\overline c$. The points of the representative cluster are independently uniformly scattered in the ball where $R$ is the radius. So the fact that BS deployment is strongly associated with user activities leads to dependence between MBSs and SBSs and dependence between the BSs and the users. In [10], it proposes a HCN model in which the MBSs and the SBSs following a PPP and an independent Matern cluster process respectively, aiming at increasing capacity in hotspots. Although the model considers the clustering property of SBSs, but doesn't take the dependence between MBSs and SBSs into consideration. Literature [11] further extends the model by using Poisson cluster process (PCP) but PCPs are independent in different tiers without considering intra-tier dependence. Moreover, nearly all works assume that the users are uniformly distributed in the whole region, so they do not consider the dependence between the users and the BSs either. Thus, the inter-tier dependence (between MBSs and SBSs) and the dependence between BSs and users have not been studied intensively. However, we know that the original purpose of HetNet is to satisfy the non-uniform user demand, the two kinds of dependence above mentioned shouldn't be neglected.

Therefore, we focus on the association and offloading in the two-tier dependent HetNet to ease the pressure of heavily loaded MBSs. The contribution of this paper can be summarized as:

\hspace{-0.35cm}1. A novel analytical two-tier HetNet model is proposed where MBSs follow a homogeneous PPP and SBSs follow a Matern cluster process whose parent points are exactly the locations of MBS. The users follow uneven distribution in the whole study region, but they are uniformly distributed in the circles centered at MBSs. Under this model, we derive the association probability and the average ergodic rate using stochastic geometry. Our difficulty lies in the distribution of desired distance between the clustered SBSs and the typical user constraint in the union of the clusters compared with the previous works. Furthermore, we obtain some interesting results by experiment evaluation.

\hspace{-0.35cm}2. On the above basis, we propose a clustering offloading scheme by deploying SBSs around the heavily loaded MBSs. We also interestingly discover that the density of MBS and the radius of the cluster can jointly control the association probabilities.

\section{System Model}\label{sec2}
The system model in this paper considers up to a two-tier deployment of the BSs. The locations of the first-tier MBSs follow a homogeneous PPP ${\Phi _m}=\left\{{{x_1},{x_2},\cdots }\right\}\subset{R^2}$ of density ${\lambda _m}$, and the locations of the second-tier SBSs follow a Matern cluster process (MCP) ${\Phi _s} = \left\{ {{y_1},{y_2},\cdots }\right\}\subset {R^2}$ whose parent point process is exactly the first-tier homogeneous PPP  ${\Phi _m}$, and the daughter points are uniformly scattered on the ball of radius $R$ centered at each parent point, assuming that the average numbers of SBS in each cluster is $\overline c$, then the density of the SBSs in the whole plane is ${\lambda _s}={\lambda _m}\overline c$. Each tier has a different transmit power ${P_i}$, $i=m$ or $s$.

For the user distribution, the users in the network are assumed to be distributed with density ${\lambda _u}$ within the circles of radius $R$ centered at each location of the MBSs and with density ${\lambda _u}'$ outside the circles ${\lambda _u} >{\lambda _u}'$. But we just focus on the users in the circles. Without loss of generality, we randomly choose a typical user located in the origin.

For the notational simplicity, we denote $k \in \left\{ {m,s} \right\}$ as the index of the tier with which a typical user associated. The downlink desired and interference signals both experience path loss, and each tier we allow different path loss exponents ${\left\{ {{\alpha _j}} \right\}_{j = m,s}}$,$\alpha  > 2$, and Rayleigh fading characterize the channel fading, i.e.,${h_{i,j}} \sim \exp (1)$. Every BS in the same tier uses the same transmit power. We thus denote ${X_k}$ as the distance between the serving BS and the typical user. We denote ${\left\{ {{D_j}} \right\}_{j = m,s}}$  as the distance of the typical user from the nearest BS in the $j$th tier. In the scenario, a user is allowed to access any tier's BSs because of open access. We consider a cell association policy based on maximum averaged biased received power(ABRP), with ${B_j}$ denoting the association bias corresponding to the $j$th tier. A user will associate with the BS that results in the highest biased averaged received signal strength. As the BSs belonging to the same tier have the same transmit power, it means a user will choose its closest MBS or SBS as its serving BS. Then we will use association probability to measure the traffic offloading.

\begin{figure}[h]
 \vspace{-0mm}
  \centering
  \includegraphics[width=3.0in]{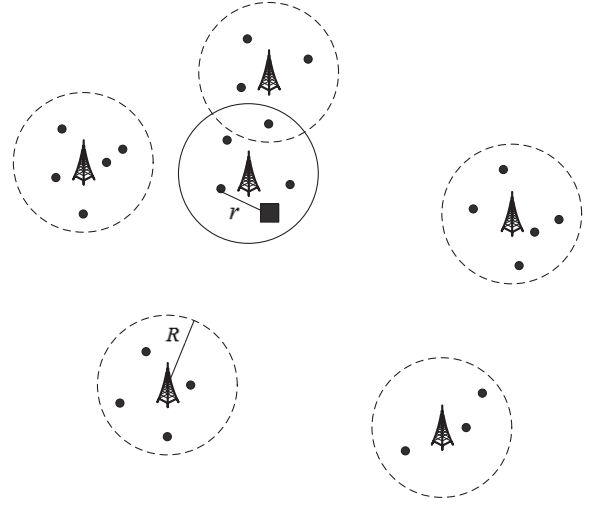}\\
  \vspace{-3mm}
 \caption{Example of the two-tier HetNet comprising a mixture of macro and small BSs: a high-power MBS is overlaid with denser and lower power SBSs (black dot). The radius of the cluster is $R$ and the black square represent the typical user.}\label{secnario}
  \vspace{-2mm}
\end{figure}

\section{Analysis Process}\label{sec3}
As mentioned above, we consider a cell association based on maximum biased-received-power, where a mobile user is associated with the strongest BS providing the highest long-term averaged biased received power at the user. The ABRP is
\begin{equation}\label{1}
 {P_{r,j}} = P{}_j{({D_j})^{ - {\alpha _j}}}{B_j}
\end{equation}
This is a long-term averaged result and fading is averaged out, so the formula (1) doesn't contain fading $h$ . However, note that the SINR model of the user associated with a BS includes fading and it will effect the distribution function of the SINR. Therefore the SINR of a typical user at a random distance $x$ from the serving BS in $k$th-tier is
\begin{equation}\nonumber
{\rm{SIN}}{{\rm{R}}_k}(x) = \frac{{{P_k}{h_k}{x^{ - {\alpha _k}}}}}{{I + {N_0}}}
\end{equation}
\begin{equation}\label{2}
I = \sum\limits_{i = m,s} {{I_i}}= \sum\limits_{i = m,s} {\sum\limits_{j \in {\Phi _i}\backslash B{S_k}} {{P_j}{h_j}{{\left| {{Y_{ji}}} \right|}^{ - {\alpha _j}}}} }
\end{equation}
Where $\left| {{Y_{ji}}} \right|$ is the distance from the BS in tier $i$ to the origin. ${N_0}$ is the thermal noise which is usually a constant and it can be neglected compared with the aggregated interference in the interference limited system.
\subsection{Distribution of the Desired Distance}\label{sec3:1}
When the location of the typical user is randomly chosen from the entire plane, the CCDF of the desired distance of an MCP was presented in [11] as
\begin{equation}\label{3}
{\rm{P}}[{D_s} > r] = \exp ( - \pi {\lambda _p}\overline c {r^2})
\end{equation}
The CCDF of the desired distance in a PPP with the density ${\lambda _m}$ is given by
\begin{equation}\label{5}
{\rm{P}}[{D_m} > r] = \exp ( - \pi {\lambda _m}{r^2})
\end{equation}
While in the model we proposed, the location of the typical user is randomly chosen from the union regions of the balls of radius $R$ centered at the parent points of the MCP. Therefore, we should calculate the CCDF of the desired distance conditioning on the event that the typical user is located within the union regions of the balls. First, the probability that the typical user is in the circles is as following based on Null Probability Theorem:
\begin{equation}\label{6}
\begin{split}
&{\rm{P}}[{D_m} \le R]\\
&\hspace{0.6cm}= 1{\rm{ - P}}[{D_m} > R] = 1 - \exp ( - \pi {\lambda _m}{R^2})
\end{split}
\end{equation}
And the conditioned CCDF of the desired distance in the first-tier PPP is
\begin{equation}\label{8}
\begin{split}
{\rm{P}}[{D_m} > r\left| {{D_m} \le R} \right.] &= \frac{{{\rm{P[r}} < {D_m} \le R{\rm{]}}}}{{{\rm{P[}}{D_m} \le R{\rm{]}}}}\\
&= \frac{{1{\rm{ - }}\exp ( - \pi {\lambda _m}({R^2} - {r^2}))}}{{1 - \exp ( - \pi {\lambda _m}{R^2})}}
\end{split}
\end{equation}
The PDF and CDF of the distance between any two points in a circle are [12]
\begin{equation}\nonumber
\begin{split}
{f_L}(l) = \frac{{2l}}{{{R^2}}}(\frac{2}{\pi }{\cos ^{ - 1}}(\frac{l}{{2R}}) - \frac{l}{{\pi R}}\sqrt {1 - \frac{{{l^2}}}{{4{R^2}}}} ),0 < l < 2R\\
\end{split}
\end{equation}
\begin{equation}\nonumber
\begin{split}
{F_L}(l) = 1 &+ \frac{2}{\pi }(\frac{{{l^2}}}{{{R^2}}} - 1){\cos ^{ - 1}}(\frac{l}{{2R}})\\
&\hspace{2.8cm}- \frac{l}{{\pi R}}(1 + \frac{{{l^2}}}{{2{R^2}}})\sqrt {1 - \frac{{{l^2}}}{{4{R^2}}}}
\end{split}
\end{equation}
The nearest distance between two points in a circle can be expressed as
\begin{equation}\nonumber
\begin{split}
{L_{\min }} = \min ({L_1},{L_2}, \cdots ,{L_{N - 1}})
\end{split}
\end{equation}
Moreover, the CDF of the minimum values of multiple independent identically distributed random variables is
\begin{equation}\nonumber
\begin{split}
{F_{{L_{\min }}}} = 1 - {[1 - {F_L}(l)]^{N - 1}}
\end{split}
\end{equation}
Then taking the derivative of $F{}_{{L_{\min }}}$, we can obtain PDF of ${L_{\min }}$
\begin{equation}\label{7}
\begin{split}
{f_{{L_{\min }}}} = (N - 1){[1 - {F_L}(l)]^{N - 2}}{f_L}(l)
\end{split}
\end{equation}
In our proposed model, there are $\overline c+1$ points scattering in a cluster uniformly. So the mapping relation is $N = \overline c  + 1,{L_{\min }} = {D_s},l = r$. Therefore, PDF of the desired distance is derived as following:
\begin{equation}\label{8}
\begin{split}
{f_{{D_m}}}(r) &= \frac{{{\rm{d\{ 1 - P[}}{D_m} > r\left| {{D_m} \le R} \right.{\rm{]\} }}}}{{{\rm{d}}r}} \\
&= \frac{{2\pi {\lambda _m}r\exp ( - \pi {\lambda _m}({R^2} - {r^2}))}}{{1 - \exp ( - \pi {\lambda _m}{R^2})}}\\
\end{split}
\end{equation}
\begin{equation}\label{9}
\begin{split}
\begin{array}{l}
{f_{{D_s}}}(r) = \overline c \frac{{2r}}{{{R^2}}}(\frac{2}{\pi }{\cos ^{ - 1}}(\frac{r}{{2R}}) - \frac{r}{{\pi R}}\sqrt {1 - \frac{{{r^2}}}{{4{R^2}}}} ) \times\\
\hspace{-0.3cm}{[\frac{r}{{\pi R}}(1 + \frac{{{r^2}}}{{2{R^2}}})\sqrt {1 - \frac{{{r^2}}}{{4{R^2}}}}-\frac{2}{\pi }(\frac{{{r^2}}}{{{R^2}}} - 1){\cos ^{ - 1}}(\frac{r}{{2R}})]^{\overline c  - 1}}
\end{array}
\end{split}
\end{equation}

\subsection{Association Probability}\label{sec3:2}
Based on our assumption,each user will connect to the BS that provides the highest ABRP.\\
\textbf{Lemma 1.}\emph{ The macro-tier association probability can be expressed as}
\begin{equation}\label{10}
\begin{split}
\begin{array}{l}
{A_m} = {\rm{P}}\left\{ {{P_m}{{({D_m})}^{ - {\alpha _m}}}{B_m} > {P_s}{{({D_s})}^{ - {\alpha _s}}}{B_s}} \right\}\\
 = {E_{{D_m}}}[{\rm{P\{ }}P{}_m{({D_m})^{ - {\alpha _m}}}{B_m} > {P_s}{({D_s})^{ - {\alpha _s}}}{B_s}\} ]\\
 = {E_{{D_m}}}[{\rm{P}}\{ {D_s} > {(\frac{{{P_m}}}{{{P_s}}} \cdot \frac{{{B_m}}}{{{B_s}}})^{ - \frac{1}{{{\alpha _s}}}}} \cdot {({D_m})^{\frac{{{\alpha _m}}}{{{\alpha _s}}}}}\} ]\\
 = \int_0^{\rm{R}} {{\rm{P}}\left\{ {{D_{\rm{s}}} > {{(\frac{{{P_m}}}{{{P_s}}} \cdot \frac{{{B_m}}}{{{B_s}}})}^{ - \frac{1}{{{\alpha _s}}}}} \cdot {r^{\frac{{{\alpha _m}}}{{{\alpha _s}}}}}} \right\} \cdot } {f_{{D_m}}}(r)dr\\
 = \frac{{2\pi {\lambda _m}}}{{1 - \exp ( - \pi {\lambda _m}{R^2})}} \times \\
\int_0^R {r\exp \{ {\rm{ - }}\pi {\lambda _p}\overline c {{(\frac{{{{\rm{P}}_m}}}{{{{\rm{P}}_s}}} \cdot \frac{{{B_m}}}{{{B_s}}})}^{ - \frac{2}{{{\alpha _s}}}}} \cdot {r^{\frac{{2{\alpha _m}}}{{{\alpha _s}}}}} - \pi {\lambda _m}({R^2} - {r^2})\} {\rm{d}}r} \\
 = \frac{{2\pi {\lambda _m}\exp ( - \pi {\lambda _m}{R^2})}}{{1 - \exp ( - \pi {\lambda _m}{R^2})}} \times \\
\int_0^R {r\exp \{ {\rm{ - }}\pi {\lambda _p}\overline c {{(\frac{{{{\rm{P}}_m}}}{{{{\rm{P}}_s}}} \cdot \frac{{{B_m}}}{{{B_s}}})}^{ - \frac{2}{{{\alpha _s}}}}} \cdot {r^{\frac{{2{\alpha _m}}}{{{\alpha _s}}}}} + \pi {\lambda _m}{r^2})\} {\rm{d}}r}
\end{array}
\end{split}
\end{equation}
{\small
\begin{equation}\label{11}
\begin{split}
{A_s} &= {\rm{P}}\left\{ {{P_s}{{({D_s})}^{ - {\alpha _s}}}{B_s} > {P_m}{{({D_m})}^{ - {\alpha _m}}}{B_m}} \right\}\\
&= \int_0^{{\rm{2R}}} {{\rm{P}}\left\{ {{D_m} > {{(\frac{{{P_s}}}{{{P_m}}} \cdot \frac{{{B_s}}}{{{B_m}}})}^{ - \frac{1}{{{\alpha _m}}}}} \cdot {r^{\frac{{{\alpha _s}}}{{{\alpha _m}}}}}} \right\} \cdot } {f_{Ds}}(r)dr\\
&= \overline c \int_0^{2R} {\exp \{ {\rm{ - }}\pi {\lambda _m}{{(\frac{{{P_s}}}{{{P_m}}} \cdot \frac{{{B_s}}}{{{B_m}}})}^{ - \frac{2}{{{\alpha _m}}}}} \cdot {r^{\frac{{2{\alpha _s}}}{{{\alpha _m}}}}}\} }\times\\
& \hspace{-0.3cm}{[  \frac{r}{{\pi R}}(1 + \frac{{{r^2}}}{{2{R^2}}})\sqrt {1 - \frac{{{r^2}}}{{4{R^2}}}}- \frac{2}{\pi }(\frac{{{r^2}}}{{{R^2}}} - 1){\cos ^{ - 1}}(\frac{r}{{2R}})]^{\overline c  - 1}}\\
& \times \frac{{2r}}{{{R^2}}}(\frac{2}{\pi }{\cos ^{ - 1}}(\frac{r}{{2R}}) - \frac{r}{{\pi R}}\sqrt {1 - \frac{{{r^2}}}{{4{R^2}}}} )dr
\end{split}
\end{equation}
}
\emph{If $\left\{ {{\alpha _m},{\alpha _s}} \right\} = \alpha $, the association probability of macro-tier and smallBS-tier is simplified to}
{\small
\begin{equation}\label{12}
\begin{split}
{A_m} = \frac{{\{ 1 - \exp [ - \pi {\lambda _m}{R^2}(\overline c {{(\frac{{{P_m}}}{{{P_s}}} \cdot \frac{{{B_m}}}{{{B_s}}})}^{ - \frac{2}{\alpha }}}{\rm{ - }}1)]\}  \cdot \exp ( - \pi {\lambda _m}{R^2})}}{{[\overline c {{(\frac{{{P_m}}}{{{P_s}}} \cdot \frac{{{B_m}}}{{{B_s}}})}^{ - \frac{2}{\alpha }}}{\rm{ - }}1] \cdot [1 - \exp ( - \pi {\lambda _m}{R^2})]}}\\
\end{split}
\end{equation}
}
{\small
\begin{equation}\label{13}
\begin{split}
\begin{array}{l}
{A_s} = \overline c \int_0^{2R} {\exp \{  - \pi {\lambda _m}{{(\frac{{{P_s}}}{{{P_m}}} \cdot \frac{{{B_s}}}{{{B_m}}})}^{ - \frac{2}{\alpha }}} \cdot {r^2}\} } \\
 \times {[\frac{r}{{\pi R}}(1 + \frac{{{r^2}}}{{2{R^2}}})\sqrt {1 - \frac{{{r^2}}}{{4{R^2}}}}-\frac{2}{\pi }(\frac{{{r^2}}}{{{R^2}}} - 1){\cos ^{ - 1}}(\frac{r}{{2R}})]^{\overline c  - 1}}\\
 \times \frac{{2r}}{{{R^2}}}(\frac{2}{\pi }{\cos ^{ - 1}}(\frac{r}{{2R}}) - \frac{r}{{\pi R}}\sqrt {1 - \frac{{{r^2}}}{{4{R^2}}}} )dr
\end{array}
\end{split}
\end{equation}
}

From Lemma 1, we observe that the density of the MBSs ${\lambda _m}$ (also the density of the parent point process ${\lambda _p}$ due to the location coincidence of the MBSs and the parent points of the MCP ) and the radius of the cluster $R$ always appear in the same form of ${\lambda _m}{R^2}$. No matter how ${\lambda _m}$ or $R$ varies, if the value of ${\lambda _m}$  maintain constant, ${A_m}$ remains unchanged as far as the typical user concerned. In the section of numerical results, we will discuss the specific relationship of these parameters. We further observe that the BS density is more dominant in determining ${A_k}$ than BS transmit power or bias factor(when $\alpha  > 2$).

The association probability of each tier is a very useful index in analyzing the network performance. It can directly represent the percentage of the users served by certain tier from the total users. So the average number of users associated with a BS in the $k$th tier is given as
\begin{equation}\label{14}
\begin{split}
{N_k} = \frac{{{A_k}{\lambda _u}}}{{{\lambda _k}}},k = m,s
\end{split}
\end{equation}
\textbf{Lemma 2.} \emph{The PDF of the distance ${X_k}$  between a typical user and its serving BS is}
\begin{equation}\label{15}
\begin{split}
{f_{{X_m}}}(x) &= \frac{{2\pi {\lambda _m}\exp ( - \pi {\lambda _m}{R^2})}}{{{A_m}(1 - \exp ( - \pi {\lambda _m}{R^2}))}}x\\
&\times \exp \{ {\rm{ - }}\pi {\lambda _p}\overline c {(\frac{{{{\rm{P}}_m}}}{{{{\rm{P}}_s}}} \cdot \frac{{{B_m}}}{{{B_s}}})^{ - \frac{2}{{{\alpha _s}}}}} \cdot {x^{\frac{{2{\alpha _m}}}{{{\alpha _s}}}}} + \pi {\lambda _m}{x^2}\}
\end{split}
\end{equation}
{\small
\begin{equation}\label{16}
\begin{split}
{f_{{{\rm{X}}_s}}}(x) &= \frac{{\overline c }}{{{A_s}}}\exp \{ {\rm{ - }}\pi {\lambda _m}{(\frac{{{P_s}}}{{{P_m}}} \cdot \frac{{{B_s}}}{{{B_m}}})^{ - \frac{2}{{{\alpha _m}}}}} \cdot {x^{\frac{{2{\alpha _s}}}{{{\alpha _m}}}}}\} \times \\
&\hspace{-1.1cm}{[\frac{x}{{\pi R}}(1 + \frac{{{x^2}}}{{2{R^2}}})\sqrt {1 - \frac{{{x^2}}}{{4{R^2}}}}-\frac{2}{\pi }(\frac{{{x^2}}}{{{R^2}}} - 1){\cos ^{ - 1}}(\frac{x}{{2R}})]^{\overline c  - 1}}\\
&\hspace{-1.0cm}\times \frac{{2x}}{{{R^2}}}(\frac{2}{\pi }{\cos ^{ - 1}}(\frac{x}{{2R}}) - \frac{x}{{\pi R}}\sqrt {1 - \frac{{{x^2}}}{{4{R^2}}}} )
\end{split}
\end{equation}
}
\emph{Proof:} We utilize the similar procedure of derivation as the Lemma 3 in [9], and the difference between the two derivation procedures is the integral upper limits. Our integral upper limits are $R$ and $2R$ corresponding to the macro-tier and smallcell-tier respectively, while in [9] it is positive infinity. So the formulas also present similar form.

\subsection{Average Ergodic Rate}\label{sec3:3}
We derive the average ergodic rate of a typical randomly located user, and it is given as [13][14]
\begin{equation}\label{17}
\Re  = \sum\limits_k^{} {{A_k}{\Re _k}} ,k = m,s
\end{equation}
We denote ${\Re _k}$ as the average ergodic rate of a typical user associated with the $k$th-tier, ${A_k}$ is the association probability of the $k$th-tier which is derived in Lemma 1. And we ignore the thermal noise in the SINR model in the following derivation.\\
\textbf{Theorem 1}. \emph{The average ergodic rates of overall network is}
\begin{equation}\label{18}
{\scriptsize
\begin{split}
&\Re = \frac{{2\pi {\lambda _m}\exp ( - \pi {\lambda _m}{R^2})}}{{(1 - \exp ( - \pi {\lambda _m}{R^2}))}}\times \\
&\hspace{-0.3cm}\int_0^R {\int_0^\infty  {x \cdot \exp \{ {\rm{ - }}\pi (\sum\limits_{j = m,s} {{x^{2/{{\hat \alpha }_j}}}{C_j}(t)}  + {\lambda _s}{{({{\hat P}_s}{{\hat B}_s})}^{2/{\alpha _s}}}{x^{2/{{\hat \alpha }_s}}} - {\lambda _m}{x^2}\}} } \\
&dtdx + \overline c \int_0^{2R} {\int_0^\infty  {\exp \{  - \pi (\sum\limits_{j = m,s} {{x^{2/{{\hat \alpha }_j}}}{C_j}(t)} } } \\
&+ {\lambda _m}{({\hat P_m}{\hat B_m})^{2/{\alpha _m}}}{x^{2/{{\hat \alpha }_m}}})\} \\
&\times {[\frac{x}{{\pi R}}(1 + \frac{{{x^2}}}{{2{R^2}}})\sqrt {1 - \frac{{{x^2}}}{{4{R^2}}}}-\frac{2}{\pi }(\frac{{{x^2}}}{{{R^2}}} - 1){\cos ^{ - 1}}(\frac{x}{{2R}})]^{\overline c  - 1}}\\
&\times \frac{{2x}}{{{R^2}}}(\frac{2}{\pi }{\cos ^{ - 1}}(\frac{x}{{2R}}) - \frac{x}{{\pi R}}\sqrt {1 - \frac{{{x^2}}}{{4{R^2}}}} )dtdx
\end{split}
}
\end{equation}
where
\begin{equation}\nonumber
\begin{split}
{\lambda _s} = {\lambda _p}\overline c ,{\lambda _p} = {\lambda _m}
\end{split}
\end{equation}
and
\begin{equation}\label{19}
{C_j}(t) = {\lambda _j}{\hat P_j}^{2/{\alpha _j}}({\hat B_j}^{2/{\alpha _j}} + Z({e^t} - 1,\alpha {}_j,{\hat B_j}))
\end{equation}
\emph{Proof: }the average ergodic rate of the macro-tier is
\begin{equation}\label{10}
\begin{split}
{\Re _m}& = {E_x}[{E_{SINR{}_m}}[\ln (1 + SIN{R_m}(x))]]\\
& = \int_0^R {{E_{SINR{}_m}}[\ln (1 + SIN{R_m}(x))] \cdot {f_{{X_m}}}(x){\rm{d}}x} \\
& = \int_0^R {\int_0^\infty  {{\rm{P}}[\ln (1 + SIN{R_m}(x)) > t]{\rm{d}}t}  \cdot {f_{{X_m}}}(x){\rm{d}}x}\\
& = \int_0^R {\int_0^\infty  {{\rm{P}}[{h_m} > ({e^t} - 1) \cdot I{P_m}^{ - 1}{x^{{\alpha _m}}}]{\rm{d}}t}  \cdot {f_{{X_m}}}(x){\rm{d}}x}\\
& = \int_0^R {\int_0^\infty  {{L_{{I_m}}}(({e^t} - 1){P_m}^{ - 1}{x^{{\alpha _m}}})} } \\
& \hspace{1.5cm}\cdot{L_{{I_s}}}{\rm{(}}({e^t} - 1){P_m}^{ - 1}{x^{{\alpha _m}}}{\rm{)d}}t \cdot {f_{{X_m}}}(x){\rm{d}}x
\end{split}
\end{equation}
With the similar method, we can obtain the as
\begin{equation}\label{20}
\begin{split}
{\Re _s}& = \int_0^{2R} {\int_0^\infty  {{L_{{I_m}}}(({e^t} - 1){P_s}^{ - 1}{x^{{\alpha _s}}})} } \\
& \hspace{1.7cm}\cdot{L_{{I_s}}}{\rm{(}}({e^t} - 1){P_s}^{ - 1}{x^{{\alpha _s}}}{\rm{)d}}t \cdot {f_{{X_s}}}(x){\rm{d}}x\\
\end{split}
\end{equation}
Where ${L_{{I_i}}}(z)$ is the laplace transform of ${I_i}$. For clarity of exposition, we define
\begin{equation}\label{21}
{\hat P_i} = \frac{{{P_i}}}{{{P_k}}},{\hat \alpha _i} = \frac{{{\alpha _i}}}{{\alpha {}_k}},{\hat B_i} = \frac{{{B_i}}}{{{B_k}}}
\end{equation}
Which respectively represent transmit power ratio, path loss exponent ratio and bias ratio of interering BS to the serving BS. And the laplace transform is
\begin{equation}\label{22}
\begin{split}
{L_{{I_i}}}((&{e^t} - 1){P_k}^{ - 1}{x^{{\alpha _k}}})\\
& = \exp \{  - \pi {\lambda _i}{{\hat P}_i}^{2/{\alpha _i}}{x^{2/{{\hat \alpha }_i}}}Z({e^t} - 1,{\alpha _i},{{\hat B}_i})\}
\end{split}
\end{equation}
with
\begin{equation}\label{23}
\begin{split}
Z({e^t}& - 1,{\alpha _i},{{\hat B}_i})\\
&= {({e^t} - 1)^{\frac{2}{{{\alpha _i}}}}}\int_{{{({{\hat B}_i}/({e^t} - 1))}^{2/{\alpha _i}}}}^\infty  {\frac{1}{{1 + {u^{{\alpha _i}/2}}}}} du
\end{split}
\end{equation}
Plugging (23) into (20) and (21), we obtain the average ergodic rate of each tier. Furthermore, plugging(10)(11)(20) and (21) into (17), we achieve the average ergodic rate of entire network.\\

\section{Numerical Results and Discussion}\label{sec4}
In Fig. 2, we obtain the average ergodic rate using Monte Carlo simulations for comparing the two-tier PPPs and our proposed hybrid model (PPP+MCP). Our simulation parameters are as follows: $({P_m},{P_s}) = (53,33)$ dBm, $\alpha  = 4, {B_m}/{B_s} = 1, {\lambda _m} = 1/(\pi {500^2})$ . It shows the average ergodic rate versus the intensities of SBS ${\lambda _s}$. The blue  line and red line are the average ergodic rate of PPPs and our proposed model, respectively. From the numerical results from the observations that for MCP, a large number of daughter nodes within each cluster achieve a higher ergodic rate than PPP because of the non-uniform distribution of users.
\begin{figure}[htbp]
  \begin{minipage}[t]{0.5\linewidth}
  \centering
  \includegraphics[width=1.7in]{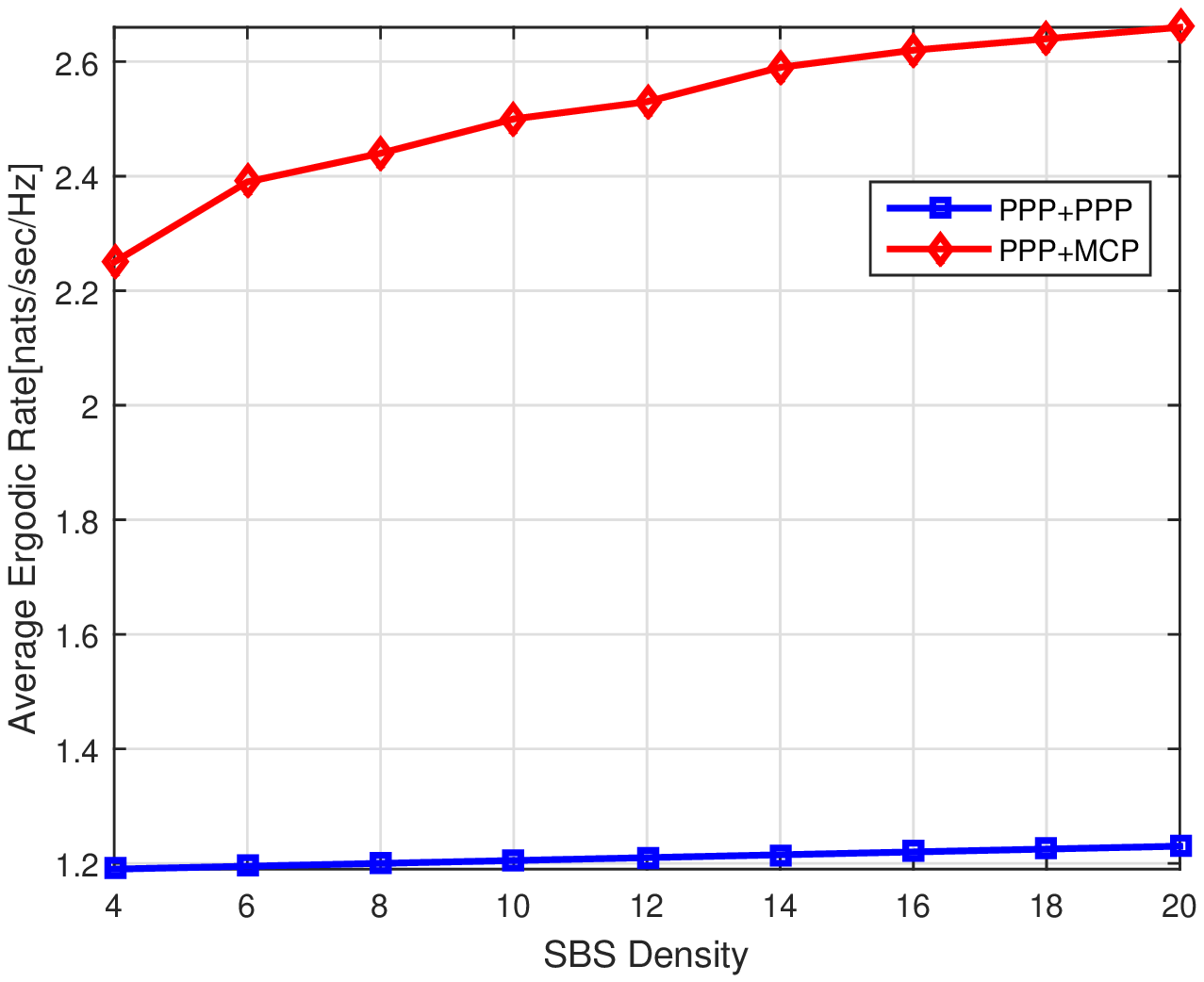}
  \caption{Average ergodic rate comparison for varying SBS density}
  \label{fig:side:a}
  \end{minipage}%
  \begin{minipage}[t]{0.5\linewidth}
  \centering
  \includegraphics[width=1.7in]{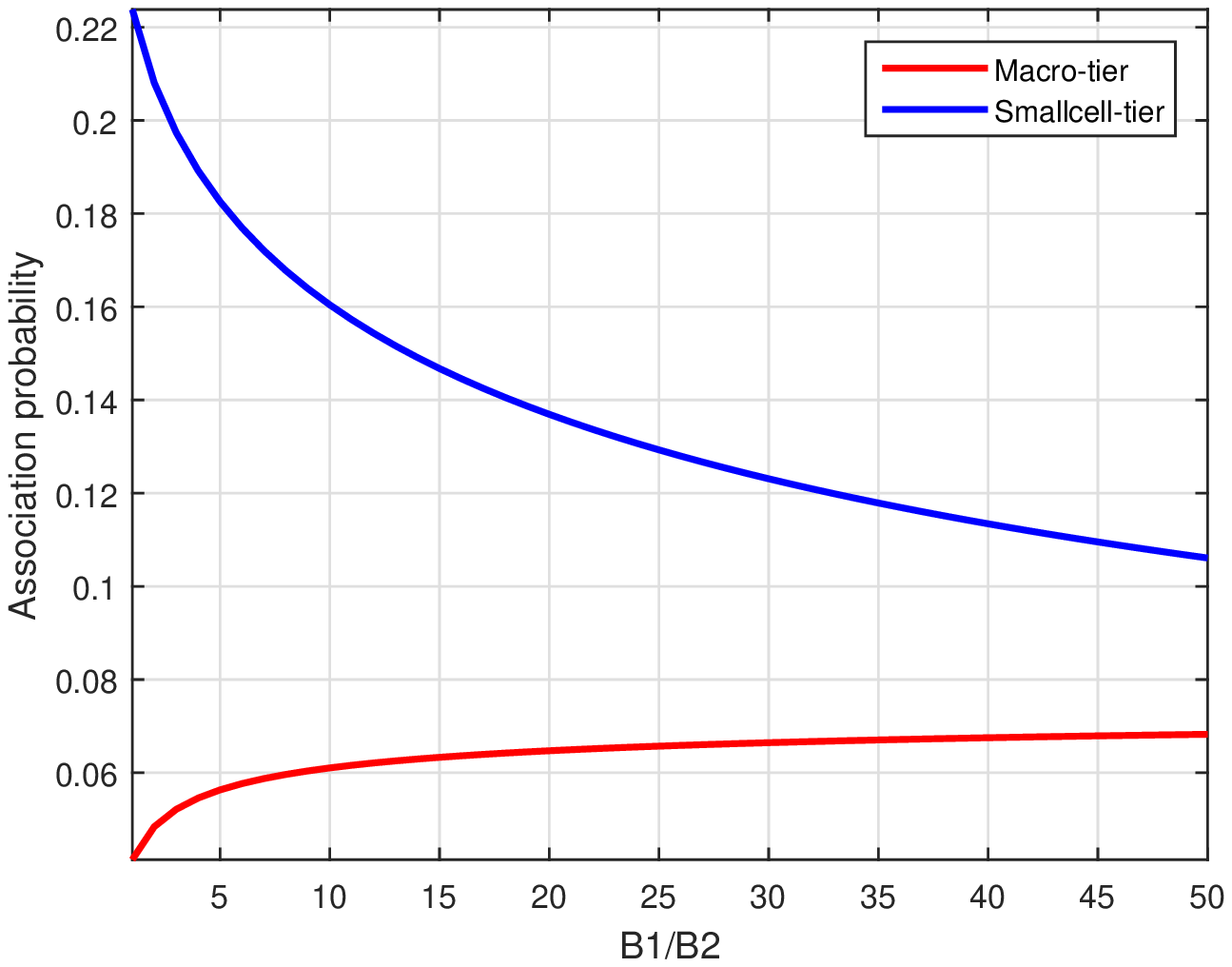}
  \caption{Effect of association bias ratio on association probability}
  \label{fig:side:b}
  \end{minipage}%

  \begin{minipage}[t]{0.5\linewidth}
  \centering
  \includegraphics[width=1.7in]{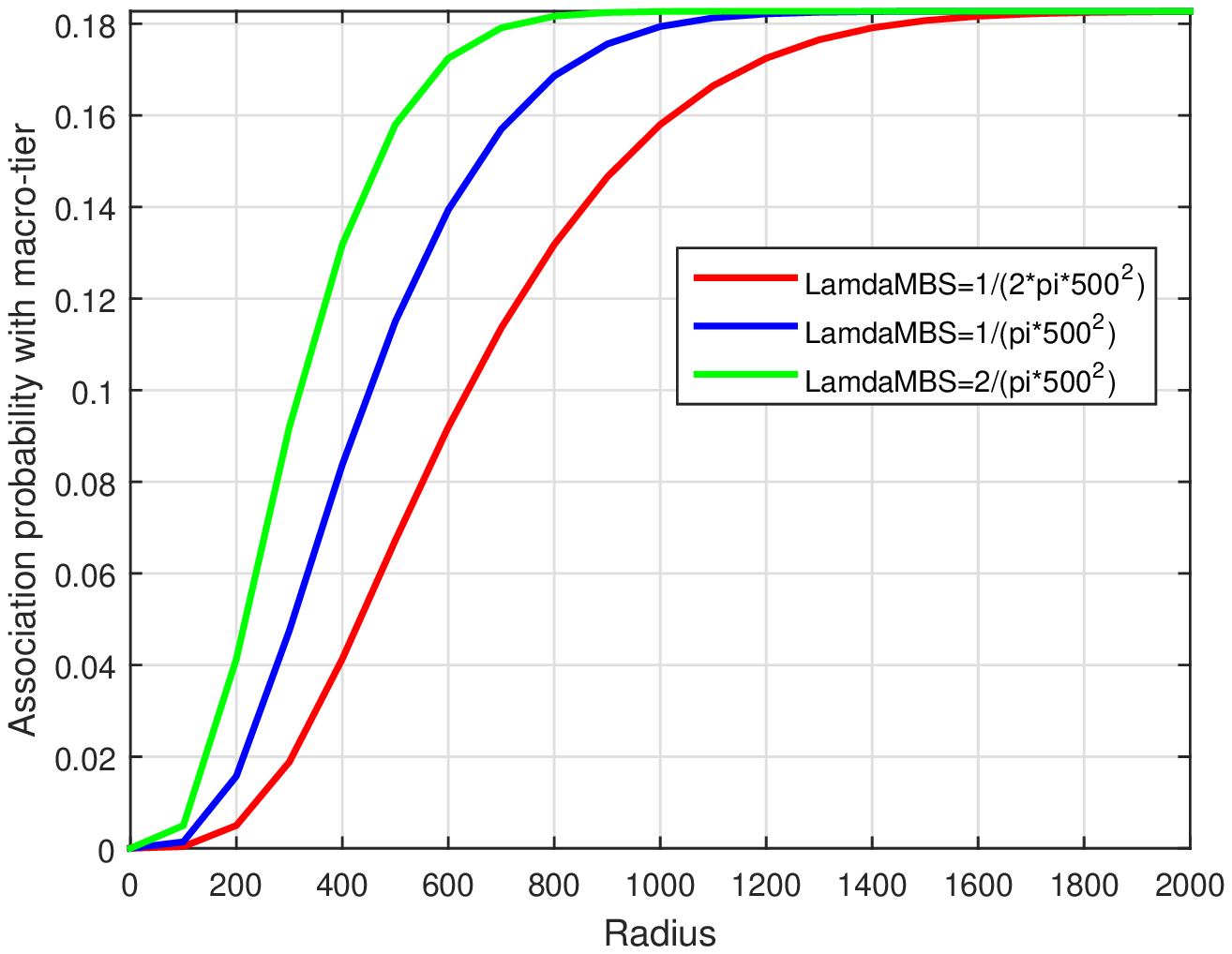}
  \caption{Effect of radius of cluster on MBS association probability}
  \label{fig:digit}
  \label{fig:side:c}
  \end{minipage}%
  \begin{minipage}[t]{0.5\linewidth}
  \centering
  \includegraphics[width=1.7in]{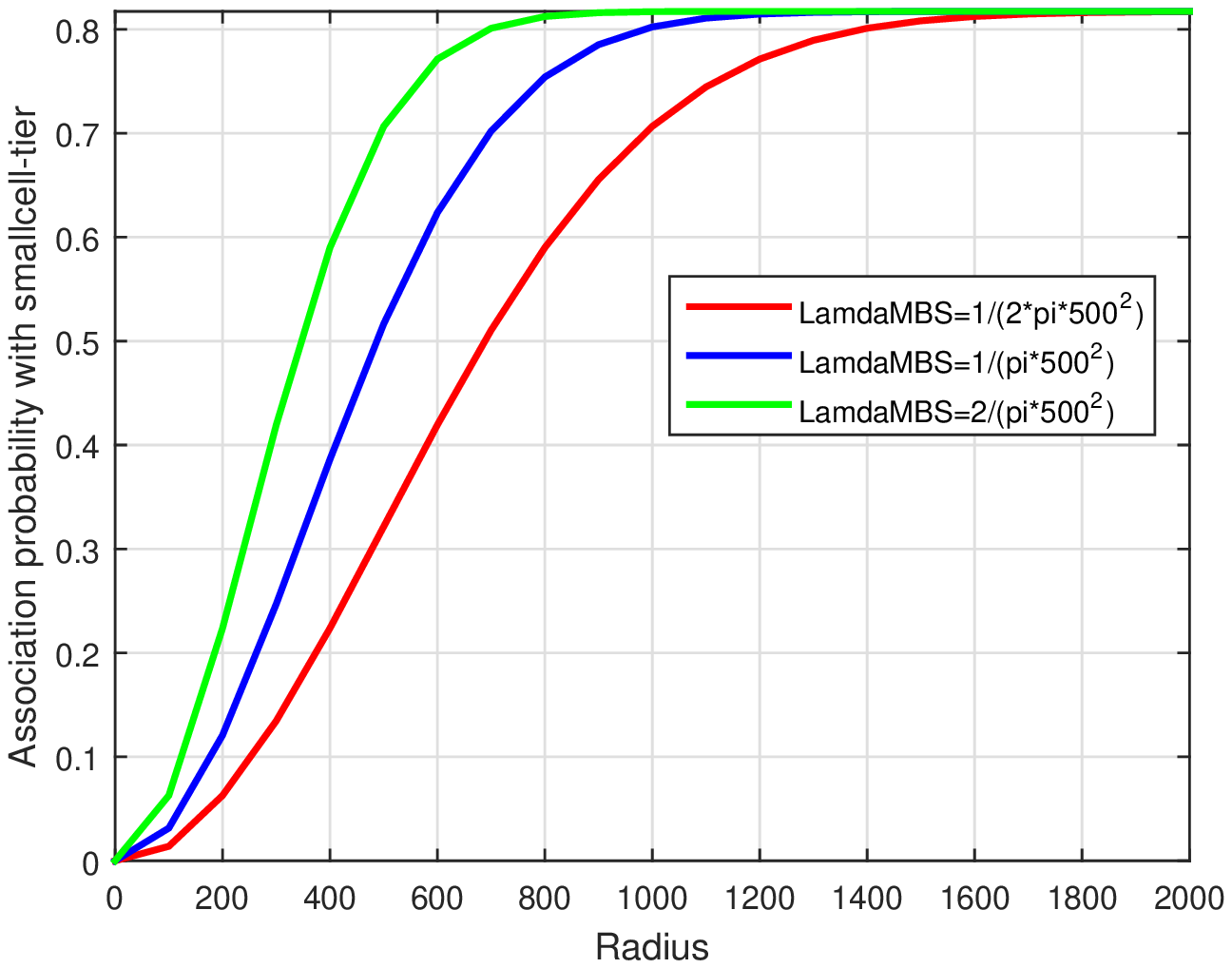}
  \caption{Effect of radius of cluster on SBS association probability}
 \label{fig:side:d}
   \end{minipage}%
\end{figure}

In Fig. 3, we explore the relation between association probability and bias ratio where the increasing ratio means the power amplification of MBS is larger than that of SBS. Higher bias ratio leads to the consequence that more user are offloaded from SBS to MBS. We can flexibly control the load of each tier by tune the biases. From the above figure, we also can see the association probability with SBS-tier is much higher than that with macro-tier. This means the typical user is more likely to connect to a SBS instead of a MBS, i.e., the users can be offloaded from MBSs to SBSs.

In Fig. 4 and Fig. 5, we can see that when the density of MBS is fixed, the association probabilities increase with increasing radius of clusters. This is because the SBSs and the users are distributed uniformly throughout the entire plane with the increasing radius. When the radius increases to a certain value, the users can achieve an equivalent uniform distribution, and the association probabilities will be constant. Moreover, they also show that the association probabilities under larger density reach a stable value at a faster speed, which validates the formula (12) in which the density of the MBS and the radius $R$ always occur in the integrated form of ${\lambda _m}{R^2}$ .
\section{Conclusion}\label{sec5}
In this paper, we presented a model considering both the inter-tier and user-BS dependence to analyze the effects of offloading in HetNet. The association probabilities and average ergodic rate were derived. An interesting result that the density of MBS and the radius of the clusters jointly affect the association probabilities  is obtained. Simulation and numerical results showed that the proposed model can aggressively offload the mobile users from MBSs by bias adjustment.
\section{Acknowledgement}\label{sec6}
The work was supported by National Nature Science Foundation of China Project (Grant No. 61471058), Hong Kong, Macao and Taiwan Science and Technology Cooperation Projects (2014DFT103202016YFE0122900), the 111 Project of China (B16006) and Beijing Training Project for The Leading Talents in S\&T (No. Z141101001514026).

\end{document}